\documentclass{PoS}

\title{The monopole mass in the random percolation gauge theory}

\ShortTitle{The monopole mass in the random percolation gauge theory}

\author{Pietro Giudice\\
        Universit\`a di Torino \& INFN sez. di Torino\\
        E-mail: \email{giudice@to.infn.it}}

\author{Ferdinando Gliozzi\\
        Universit\`a di Torino \& INFN sez. di Torino\\
        E-mail: \email{gliozzi@to.infn.it}}

\author{\speaker{Stefano Lottini} \\
        Universit\`a di Torino \& INFN sez. di Torino\\
        E-mail: \email{lottini@to.infn.it}}

\abstract{We study the behaviour of the monopole at finite temperature in the $(2+1)$-dimensional lattice gauge theory dual to the percolation model; by exploiting the correspondences to statistical systems, we possess powerful tools to evaluate the monopole mass both above and below the critical temperature with high-precision Monte Carlo simulations.}

\FullConference{The XXVI International Symposium on Lattice Field Theory \\
                 July 14 - 19, 2008\\
                 Williamsburg, Virginia, USA}

\newcommand{\avg}[1]{\langle #1 \rangle}
\newcommand{\eat}[1]{{}}

\begin{document}
\section{Introduction: monopoles and confinement}
	Confinement phenomenology is well described by the dual superconductor picture, proposed decades ago by Polyakov, 't Hooft and Mandelstam. The interquark potential gets its linearly increasing behaviour from the fact that the monopole condensate acquires a nonzero expectation value, squeezing the chromoelectric flux between sources in a string-like tube, as a dual version of the Meissner effect in type II superconductivity. The QCD vacuum is viewed as populated by a dense graph of magnetic flux lines, which melts down to finite-size remnants at the deconfinement point \cite{c_z}.

	According to this point of view, confinement is due to the disordering of the gauge configuration by percolation of magnetic lines in the system, and monopoles should be observed; the magnetic degrees of freedom, on the other hand, cannot simply disappear at the deconfining transition, thus monopoles should be observed also above the critical temperature, where there is no more any magnetic condensate (i.~e.~the \emph{disorder parameter} $\avg{\phi_m}$ drops to zero).

	A discrete Abelian gauge theory (as the one presented in Sec.~\ref{sec:percmonopoles}) cannot exhibit dynamical monopoles: they have to be inserted as external static sources, along with their Dirac string transporting the excess flux to infinity. This is forcefully realised by means of a nonlocal operator; as we will see, in the dual formulation of gauge theories, locality is restored.

	This paper, which will present a numerical study of the behaviour of the monopole condensate and the monopole mass(es) in a lattice gauge theory, is structured as follows: the next Section gives a quick overview on the theory we examined and motivates the emergence of signals from monopoles in such a setup, Sec.~\ref{sec:mc} describes in detail the algorithm used and the data generated by Monte Carlo simulation, and finally in Sec.~\ref{sec:results} we present the results obtained and draw some general conclusions.

	Before proceeding, a couple of remarks are in order. The monopole mechanism is well understood in Abelian gauge theories (such as the one we will inspect); hovewer, the interest mostly regards non-Abelian theories: anyway, since the Abelian degrees of freedom seem to encode the relevant physics, one safely resorts to techniques such as the Abelian projection, thus recovering the physical relevance of very simple gauge groups. Secondly, much of the literature about monopole mass in gauge theories deal with a \emph{monopole density} operator $\rho(\mathbf{x})$, so to measure their statistical behaviour (for instance, \cite{genova07}): this is different from using a quantum monopole creation operator, as was done e.~g.~in \cite{digiacomo_etal_2} (and more specifically in \cite{luca}), and that is what percolation allows to perform.
\section{Percolation and monopoles}
	\label{sec:percmonopoles}
	The numerical investigation was carried on in a particular $(2+1)$-dimensional pure gauge model, namely the dual of the random percolation statistical system. This theory, first formulated in \cite{rpagt}, was proven to encode all features of a confining gauge theory despite its apparently trivial formulation; moreover, its simplicity opens the way to very efficient simulation techniques: a high numerical accuracy can therefore be attained in an affordable time.

	The basic idea is to extend to the $Q=1$ case the duality relation existing between every discrete gauge theory with the symmetric $Q$-element group $S_Q$ and the corresponding $Q$-state Potts model. In this view, the Ising model is viewed as a representative of the $Q=2$ case. The partition function of the spin model can be recast in terms of a sum over all possible partitions of the set of sites into aligned connected components (Fortuin-Kasteleyn clusters), whose state (among the $Q$ available ones) is still undetermined. Since the F-K formulation is explicitly analytical in $Q$, one can reach the $Q=1$ case in a smooth way, thus arriving at the random percolation model, where the sites carry no degrees of freedom and the links are simply turned \textit{on} or \textit{off} according to a probability $p$, which is the system's coupling constant.

	The recipe for evaluating a loop observable in the spin dual of the $S_Q$ gauge theory, which involves the winding number, modulo $Q$, of the F-K clusters around the loop contour, can be naturally adapted to the $Q=1$ case in a simple way: a Wilson loop is zero if there are \textit{on} clusters topologically linked to its border, one otherwise. Starting from this definition, several numerical as well as analytical consequences have been obtained (see for example \cite{pietro} at this Conference), proving the status of well-behaving gauge theory to the model under study, even though a direct formulation is (still) unknown and its properties can be investigated only from the dual formulation. The above considerations also lead to identifying the on links, in the lattice dual to the gauge one, with the magnetic flux lines of the field-theoretic approach.

	Random percolation systems undergo a continuous phase transition at a critical value $p_c$, corresponding to the sudden appearance of an infinite connected cluster (responsible for the area-law damping of large Wilson loops); this can be mapped to a finite-temperature deconfinement transition since the value of $p_c$ is a function of the temperature, thought of as the inverse of the imaginary time compactification length $\ell$ of the lattice. One can then work at a fixed probability $p_c(T_c)$ and vary the temperature above and below the transition temperature by shrinking or enlarging $\ell$. Moreover, a suitable change of both $T_c$ and $p_c(T_c)$ can keep the system along a constant physics trajectory, allowing to check for regularisation-independence of the results.

	In a gauge theory on the lattice, a magnetic monopole in $\mathbf{x}$ translates to the elementary cube dual to $\mathbf{x}$ having a nonzero total outgoing flux. Although the following reasoning is valid for all values of $Q$, we will focus on the Ising case for simplicity (in which, by the way, monopole and antimonopole coincide). To insert a static monopole, along with its Dirac string, in the gauge model, one has to flip the sign of the coupling, i.~e.~$\beta\to-\beta$, for an infinite stack of plaquettes dual to a line connecting $\mathbf{x}$ to infinity.

	Since the Kramers-Wannier duality maps the flip of the plaquette $\Box$ to the operator $\sigma_x \sigma_{x+1}$, involving the two spin variables located at the endpoints of the link dual to $\Box$, it follows that the above procedure for the static external monopole results in just the one-spin operator $\sigma_x$.

	As for the monopole-antimonopole correlator, by exploiting the freedom of displacing the nonphysical flux lines, and thanks to the fact that two superimposed flips cancel out, we can express any monopole two-point function to a spin-spin correlation function in the Ising model.

	We now resort to the Fortuin-Kasteleyn approach to derive an operative prescription for the monopole correlation which is valid also in the $Q=1$ case. In the F-K formulation, every connected component of the configuration can still be assigned any of the $Q$ available states: as a result, by averaging over all possible state assignments of a given cluster configuration, we have the following simple result: $\sigma_x\sigma_y=1$ if they belong to the same component, otherwise $\sigma_x\sigma_y=0$. For the percolation case, then, we define the correlation function $C(\mathbf{x},\mathbf{y})$ as an operator that measures whether the two points belong to a single cluster or not.
\section{Monte Carlo approach}
	\label{sec:mc}
	\subsection{Aim of the investigation}
		We performed a study of the monopole mass and condensate in the $(2+1)$-dimensional random percolation gauge theory by means of Monte Carlo simulation. More precisely, the monopole mass is extracted from the exponential decay of the zero-momentum projected correlation function (the index $1$ represents a spatial axis of the system)
			\begin{equation}
					\label{eq:define_cr}
				C(r) \equiv \sum_{\mathbf{y}_1=\mathbf{x}_1+r} C(\mathbf{x},\mathbf{y})\;\;,
			\end{equation}
		while the monopole condensate $\phi_m$, in view of the consideration presented in Sec.~\ref{sec:percmonopoles}, becomes the magnetisation operator $\avg{\sigma}$ and is thus to be identified with the \emph{strength of the infinite cluster} $\avg{s}$, that is the fraction of nodes belonging to the largest connected component in the percolative configuration. This quantity will remain finite in the thermodynamic limit only in the confined phase, i.~e.~when $p>p_c$, allowing for the formation of a component spanning the entire system extent.

		The function $C(r)$ will exhibit an exponentially decreasing behaviour governed by the monopole mass $M$,
			\begin{equation}
				C(r) = A e^{-M r} + K \;\;,
			\end{equation}
		with an asymptotic value which vanishes in the deconfined phase ($K=0$), and in the confined phase coincides with the square of the monopole condensate: $K=\avg{\phi_m}^2$. The possibility to have more than one massive monopole state cannot be ruled out; in that case, a multiple-exponential would be found (it is understood $M_1<M_2<\ldots$):
			\begin{equation}
				C(r) = A_1 e^{-M_1 r} + A_2 e^{-M_2 r} + \ldots + K \;\;.
			\end{equation}
	\subsection{Algorithm and simulation parameters}
		We consider a $L^2\times\ell$ lattice, $L$ being the spatial side which goes to infinity in the thermodynamic limit and $\ell=1/(aT)$ the imaginary time extent ($a$ is the lattice spacing). We set periodic boundary conditions in all directions, and work at various choices of the occupation probability $p_c$, corresponding to as many values for the critical temperature $T_c(p_c)$. We probe the connection correlator illustrated above for a variety of temperatures, above and below the transition temperature, keeping in mind that the physical quantity useful for comparing different settings is $T/T_c$. We also measure independently the condensate, that is $\avg{s}$, in order to avoid large systematics in estimating, in the confined phase, the background in the correlators.

		To reach satisfactory statistics, for each lattice geometry and occupation probability an amount of configurations is inspected varying from $100$ and $900$ thousands. We soon found out that the presence of a nonzero background poses serious threats on the accuracy of the results, mainly because of autocorrelation issues, so we generated data using two different approaches in the two phases: at $T>T_c$ we measured all spatial distances $r=L/4,\ldots,L/2$ in a single sweep, while for $T\leq T_c$ we performed 8 different runs to independently measure the correlators for $r=1,9,\ldots$, $r=2,10,\ldots$ and so on up to $r=8,16,\ldots,L/2$. Also the analysis did proceed in a different way, depending on the approach which gave the most stable results: in the confined phase we relied on standard fits with the sliding window technique, that is, by trying all possible $[r_\mathrm{min}:r_\mathrm{max}]$ fit intervals and looking for stable plateaux, while in the deconfined phase we compared the quantity $C(r+1)/C(r)$ obtained from numerical data with the predicted values by varying $M$ (amplitudes do cancel out in the ratio).

		Once we have computed the masses and the background for an array of spatial extents $\{L\}$, we proceed with a finite-size-scaling analysis to reach the thermodynamic limit. We studied most intensely the range $[T_c/2:T_c]$ in the confined phase, while in the deconfined phase we went up to $\sim 8 T_c$. All simulation settings are listed in Table \ref{tab:simsettings}.
		\begin{table}
			\begin{center}
			\begin{tabular}{|c|c|c|c|}
				\hline
					$1/T_c$ & $p_c$ & $1/T$, deconfined & $1/T$, confined \\
				\hline
					$7$ & $0.268459$ & $1,2,3,4,5,6$ & - \\
					$8$ & $0.265615$ & $1,2,3,4,5,6,7$ & $8,9,10,11,12,13,14,15,16,17; 48$ \\
					$9$ & $0.263658$ & $7,8$ & $9,10,11,12$ \\
					$11$ & $0.260620$ & $9,10$ & $11,12,13$ \\
					$13$ & $0.258571$ & $11,12$ & $13,14,15$ \\
				\hline
				\hline
					$1/T_c$ & $p_c$ & \multicolumn{2}{|l|}{$1/T$, background only:} \\
				\hline
					$15$ & $0.257101$ & \multicolumn{2}{|c|}{$16,17,18$} \\
					$17$ & $0.255998$ & \multicolumn{2}{|c|}{$18,19,20$} \\
					$19$ & $0.255142$ & \multicolumn{2}{|c|}{$20,21,22,23$} \\
				\hline
			\end{tabular}
			\caption{System geometries and occupation probabilities explored in this investigation. Spatial sides range from $L=64$ to $L=512$ for monopole mass extraction and from $L=312$ to $L=650$ for the background, except the case $1/T_c=19, 1/T=20$, where we went up to $L=1000$. All inverse temperature are expressed in lattice spacings.}
			\label{tab:simsettings}
			\end{center}
		\end{table}
		The algorithm goes as follows: a new configuration is created from scratch, then a standard Hoshen-Kopelman cluster reconstruction is performed, so that each node possesses a label denoting the index of the connected component it belongs to. Now, since what we are interested in is the zero-momentum projection of the correlator, each temporal plane of the lattice is conveniently assigned a (sparse) vector listing how many nodes it contains for each cluster in the system: thus, the evaluation of $C(r)$ consists simply of scalar products between pairs of such vectors (this results in an increase in the speed of the algorithm of order $\sim L$ with respect to actually following Eq.~\ref{eq:define_cr}).
\section{Results and conclusions}
	\label{sec:results}
	\subsection{Deconfined phase}
	In the deconfined phase of the theory, for each finite spatial size, the correlator data show an effective background; anyway, its value smoothly vanishes in the $L \to \infty$ limit as expected. Nevertheless, to extract the masses it is mandatory to subtract the background value from $C(r)$. In this phase the functional form of the correlator is nicely described by a single-exponential behaviour; actually, due to the periodic b.~c., we included the reversed term and used the function
		\begin{equation}
			C(r) = A [e^{-M r}+e^{-M (L-r)}]\;\;.
		\end{equation}
	The finite-size scaling of the masses with $L$ did give good results and allowed for a very precise determination of their thermodynamic limit via the relation $M_L=M_\infty+aL^{-1/\nu}$, $\nu=4/3$ being the thermal exponent for two-dimensional percolation.
	\begin{figure}
	\begin{center}
		\includegraphics[height=6.7cm,angle=270]{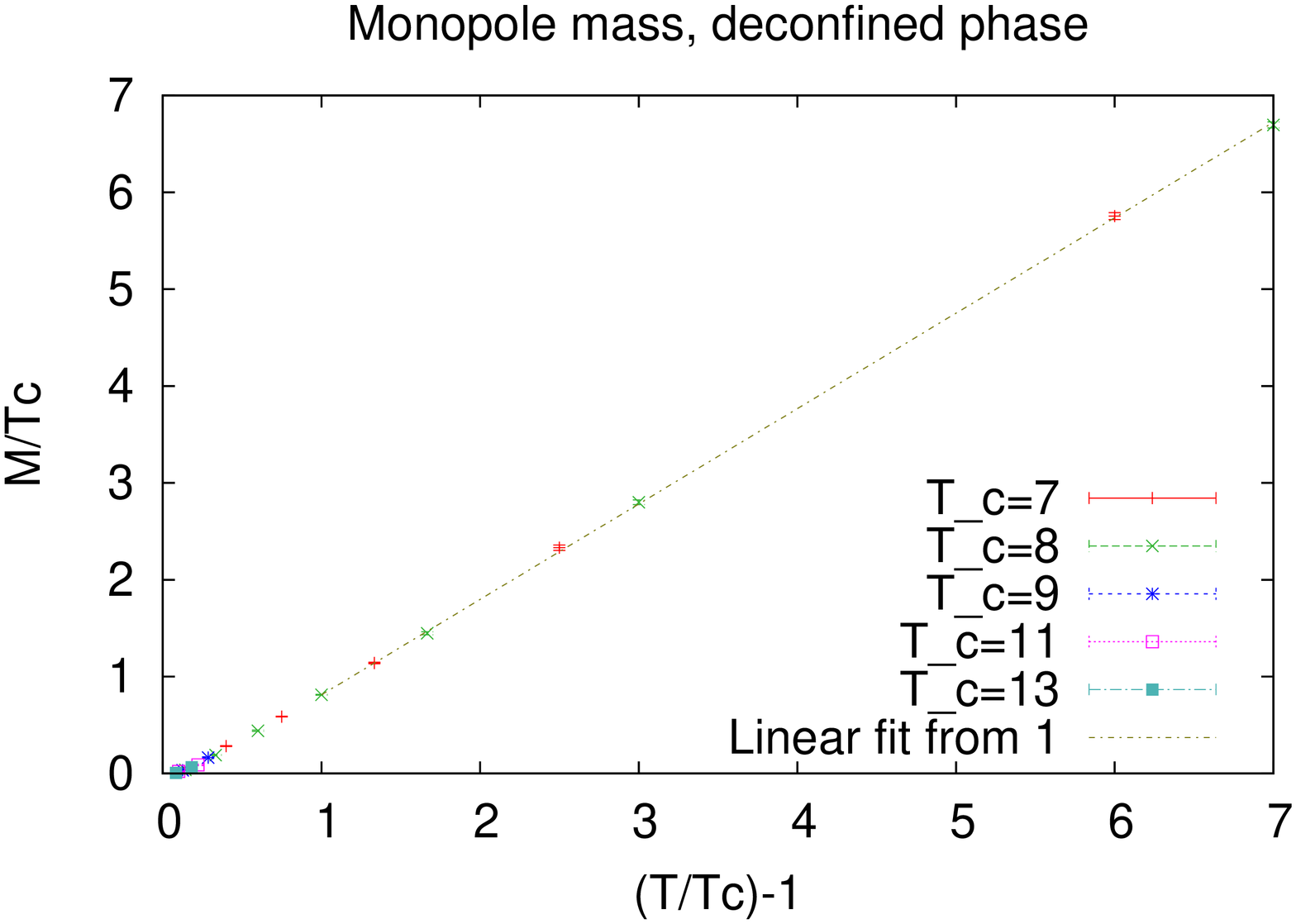}\includegraphics[height=6.7cm,angle=270]{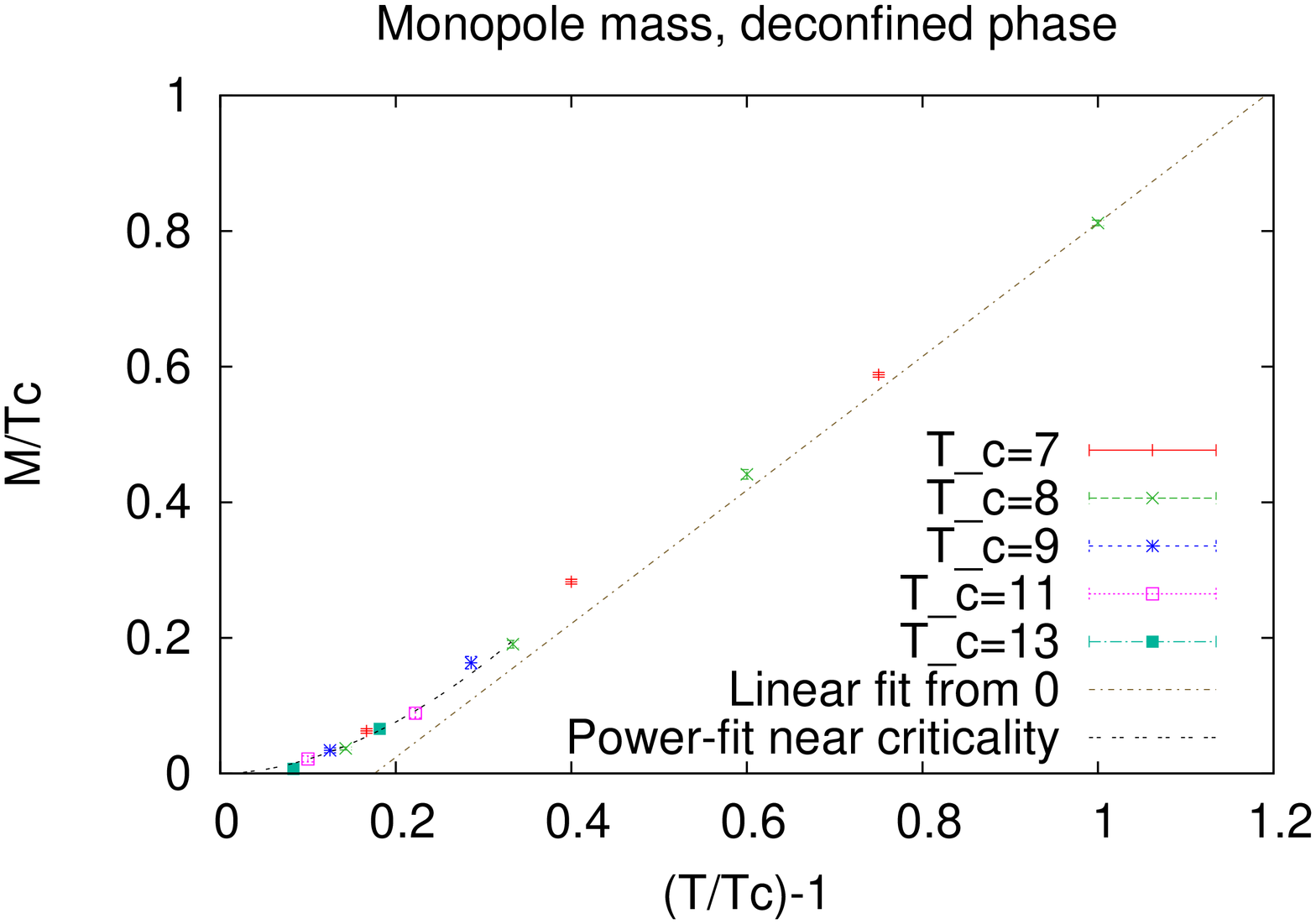}
		\caption{Mass of the only monopole in the deconfined phase, in units of $T_c$, as a function of $(T/T_c)-1$. The linear behaviour is clearly visible up to $8$ times the transition temperature. On the right, the scaling of the numerical data for $m$ near criticality along with the estimated power-law with exponent $a=1.851(53)$.}
		\label{fig:deconf}
	\end{center}
	\end{figure}
	The resulting mass $M$ behaves linearly in the adimensional variable $t=\frac{T-T_c}{T_c}$ up to $\sim8T_c$, in agreement with the expectations, being $T$ the only physical quantity involved far enough from criticality. The straight line, however, does not start exactly at $T_c$, but rather at about $T\sim1.18 T_c$: in the narrow region just above the transition we found instead a power-law behaviour $M/T_C\sim T^a$ (Fig.~\ref{fig:deconf}).
	\subsection{Behaviour at the transition and confined phase}
	At the critical point, both the single-mass and the background from the data yield a zero value in the $L\to\infty$ limit, in agreement with the expectation, with exponents, from percolation theory, respectively $1/\nu$ and $\eta=\frac{5}{24}$.

	In the confined phase, the data suffered from greater systematics due to the dominance of the background over the signal. After subtracting the background from the correlator, we found clear evidences of the existence of \emph{two} massive states, each with its own behaviour in the interval $[0;T_c]$. The fundamental mass $M_1$ follows a linear decreasing while approaching criticality, and again the intercept ($\sim 0.9518(2)$) hints at a power-law vanishing of the mass with $T$ near deconfinement; unfortunately, it was not possible to measure the exponent on this side of the transition.

\begin{figure}
	\begin{center}
		\includegraphics[height=7.3cm,angle=270]{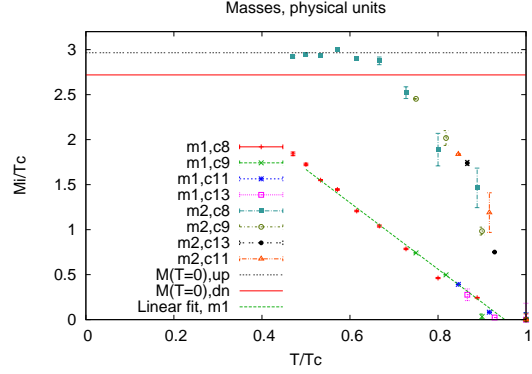}
		\caption{The two masses $M_1/T_c, M	_2/T_c$ of monopoles in the confined phase as a function of $T/T_c$. The line is a linear fit for the lightests mass; the second one rapidly reaches the zero-temperature value (whose compatibility range is delimited by the two horizontal lines).}
		\label{fig:confined}
	\end{center}
	\end{figure}

	The second mass $M_2$ instead rises abruptly from $T_c$, reaching a constant value as early as $T\sim0.65T_c$. This value is compatible with the zero-temperature mass $M_{T=0}$: since at $T=0$ we found a single-mass signal again, we conclude that $M_1$ and $M_2$ are degenerate both at $T=T_c$ (where they are zero) and at $T=0$, though the intermediate behaviour is very different for the two states. The zero-temperature value coincides, as expected, with the mass of the lightest glueball in the scalar positive-parity $0^+$ (Fig.~\ref{fig:confined}).
	The fact that the curves obtained for different choices of $(T_c,p_c)$ fall onto each other if we use the physical variables $(M_i/T_c,T/T_c)$ can be advocated as a proof of the regularisation-independence of the above results.
	\begin{figure}
		\begin{center}
		\includegraphics[height=7.2cm,angle=270]{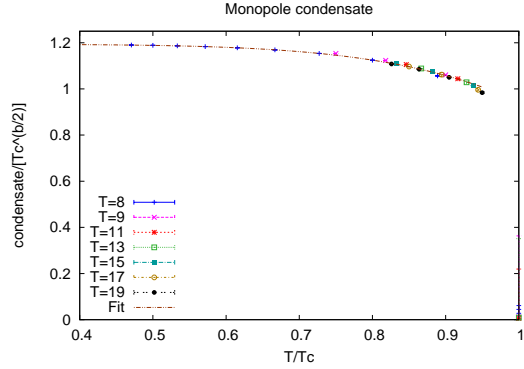} 
		\caption{The value of the monopole condensate as a function of the relative temperature $T/T_c$, after the rescaling of the curves as indicated in the text. The line is a single-exponential fit to data with exponent $q=5.8(2)$.}
		\label{fig:fondo}
		\end{center}
	\end{figure}
	As for the monopole condensate $\avg{\phi_m}$, the curves from different regularisations fall onto each other with the physical rescaling $\avg{\phi_m} \mapsto \frac{\avg{\phi_m}}{T_c^{b/2}}$, with $b = 0.874(16)$ compatible with the 3-dimensional critical index $\nu_{3d}\simeq0.8765$ of percolation. The departure from the zero-temperature value for the background follows a single power-law behaviour up to at least $T\sim0.9T_c$ with exponent $q=5.8(2)$ (Fig.~\ref{fig:fondo}). It would be interesting to compare with the analogue quantities for other theories, such as the $Z_2$ gauge model, to check whether the dimension of the monopole operator, near the deconfining point, is universally equal to $(\mbox{length})^{-\nu/2}$.

\end{document}